\documentclass[aps,prb,twocolumn,showpacs,groupedaddress]{revtex4}

\usepackage{amssymb}   

\usepackage{graphicx}
\usepackage{dcolumn}
\usepackage{bm}
\usepackage{amssymb} 

\begin{document}

\title{Crystalline electric fields and the magnetic ground state of the novel Heusler intermetallic YbRh$_{2}$Pb}

\author{D. A. Sokolov$^{1}$$\dag$,  M. S. Kim$^{1}$, M. C. Aronson$^{1,3}$, C. Henderson$^{2}$, and P. W. Stephens$^{3}$}

\address{$^{1)}$ Department of Physics, The University of Michigan, Ann Arbor, Michigan 48109-1120, USA\\
}
\address{$^{2)}$ Department of Geological Sciences, The University of Michigan, Ann Arbor, Michigan 48109-1120, USA\\
}
\address{$^{3)}$ Department of Physics and Astronomy, State University of New York, Stony Brook, New York 11794, USA\\
}

\begin{abstract}
{We have synthesized a new intermetallic compound with a distorted
Heusler structure, YbRh$_{2}$Pb. We present a study of the magnetic,
thermal, and transport properties. Heat capacity measurements
revealed that YbRh$_{2}$Pb orders magnetically below T$_{N}$=0.57 K
from a paramagnetic state with substantial crystal electric field
splitting. Magnetic field further splits the ground state, which
leads to the suppression of magnetic order in YbRh$_{2}$Pb.}

\end{abstract}
 \maketitle

\section{INTRODUCTION}
Analysis of the magnetic properties of compounds containing
rare-earths is complicated when the Kondo temperature and
crystalline electric field (CEF) splitting of the ground state are
comparable to the ordering temperature. The CEF reduces the
degeneracy of the total angular momentum $\emph{J}$, often leading
to deviations from the Curie-Weiss behavior of the temperature
dependent magnetic susceptibility. The CEF scheme in high symmetry
crystal structures can be established using heat capacity and
magnetic susceptibility measurements, making it straightforward in
this case to isolate possible Kondo effects as well as phenomena
related to the magnetic ordering. The Heusler compounds are an
especially attractive class of materials in which to pursue such
studies, as they contain a cubic lattice of rare-earth
ions~\cite{webster1969}. Rare-earth based Heusler compounds
generally order antiferromagnetically at low temperatures due to the
conduction electron mediated Ruderman-Kittel-Kasuya-Yosida (RKKY)
interaction among well-localized rare earth moments
~\cite{ishikawa1982,malik1985,kierstead1985,shelton1986,stanley1987,li1989,
seaman1996,aoki2000,gofryk2005}. Intriguingly, coexistence of
antiferromagnetism and superconductivity has been reported for some
of these compounds ~\cite{ishikawa1982,kierstead1985,shelton1986,
li1989,seaman1996,aoki2000}. In this article we report a new Yb
based tetragonally distorted Heusler compound,YbRh$_{2}$Pb, with a
magnetic ordering temperature comparable to the splitting in the
ground state manifold. The tetragonal distortion of the cubic
Heusler structures was reported previously and was attributed to an
electronic instability of the band Jahn-Teller type\cite{suits1976}.
We demonstrate that the degeneracy of the manifold in YbRh$_{2}$Pb
is partially lifted by both the CEF and the magnetic field. The
magnetism of YbRh$_{2}$Pb, characterized by total angular momentum,
$\emph{J}$, is suppressed by the reduced degeneracy of the ground
state.

\section{EXPERIMENTAL DETAILS}
Samples of YbRh$_{2}$Pb were grown from Pb flux. They had the
appearance of faceted cubes with typical dimensions of 2 mm. The
grains were etched to remove excess Pb using a 1:1 solution of
H$_{2}$O$_{2}$ and acetic acid. A small amount of one grain was
crushed, and a powder X-ray diffraction pattern was collected at
room temperature. A Si(111) double crystal monochromator selected a
beam of 0.70050(2) {$\AA$} X-rays at the X16C beamline of the
National Synchrotron Light Source at Brookhaven National Laboratory.
The diffracted X-rays were analyzed by a Ge(111) crystal and
detected with a NaI scintillation counter. The sample, on a quartz
zero-background holder, was oscillated 2 degrees at each point
during data collection. Microanalysis measurements were performed on
a CAMECA SX100 electron microprobe at the University of Michigan.
Electron beam voltage was 20 keV, beam current was 10 nA, and the Yb
L$\alpha$, Rh L$\alpha$, and Pb M$\alpha$ X-ray intensities were
calibrated from synthetic YbPO$_{4}$, synthetic CeRhSn, and natural
PbS standards respectively. Matrix elements were calculated using
CAMECA PAP data reduction routine. The magnetic susceptibility was
measured using a Quantum Design SQUID magnetometer at temperatures
from 1.8 K to 300 K. Measurements of the electrical resistivity and
heat capacity were performed using a Quantum Design Physical
Property Measurement System at temperatures from 0.35 K to 300 K and
in magnetic fields up to 4 T.

\section{RESULTS AND DISCUSSION}
Electron microprobe experiments found that the stoichiometry was
uniform over the surface of the samples, with the elemental ratios
for Yb:Rh:Pb of 1$\pm0.02:2\pm0.04:1\pm$0.02. Examination of several
small grains chipped from the faceted as-grown samples with a Bruker
Smart CCD X-ray diffractometer revealed that they were multiply
twinned. The powder X-ray diffraction pattern was indexed and
refined using Topas software\cite{topas}. The powder X-ray
diffraction pattern showed the presence of Pb and several other weak
unidentified peaks. The unit cell was found to be tetragonal, with
dimensions a=4.5235(4)$\AA$, c=6.9864(6)$\AA$, probable space group
$\emph{I4/mmm}$. The small, high symmetry unit cell suggested a
distorted Heusler alloy structure, and indeed a satisfactory
refinement was found with Yb at (0,0,0), Pb at (0,0,1/2), Rh at
(0,1/2,1/4). Refinement of occupancies showed no significant (less
than 5$\%$) mixing of atoms in the various sites. The crystal
structure of YbRh$_{2}$Pb is shown in Fig.~1. The observation of
cube-shaped grains containing multiple twins suggested that the
sample material was a cubic phase when it solidified from the melt,
and transformed into the tetragonal phase as it cooled. Similar
behavior in Heusler intermetallic phases has been noted previously,
and ascribed to an electronic band-driven Jahn-Teller
distortion\cite{suits1976}.

\begin{figure}
\includegraphics[scale=0.45]{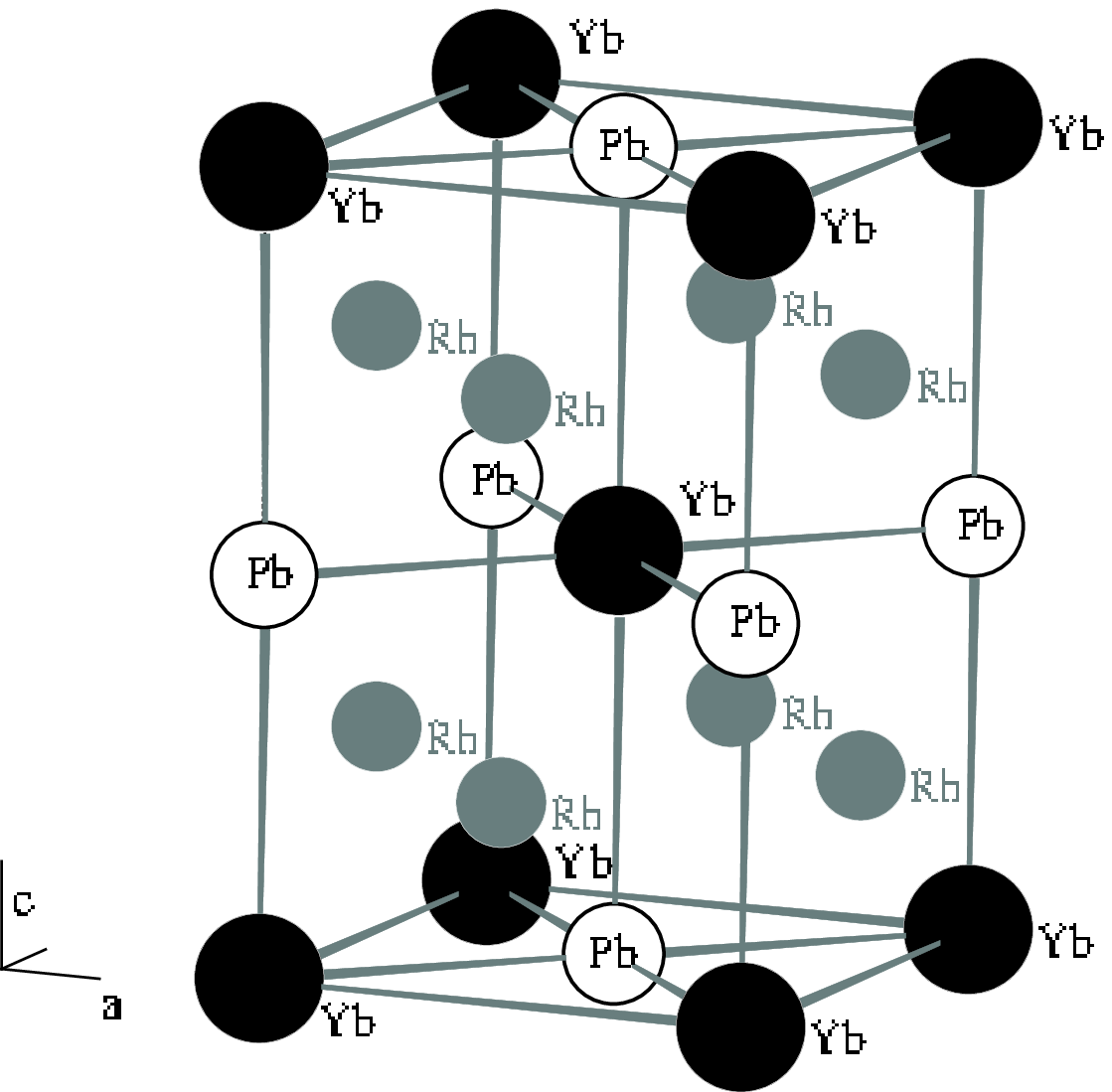}
\caption{\label{fig:epsart} Crystal structure of YbRh$_{2}$Pb. Rh
atoms occupy (0,1/2,1/4) sites, Yb atoms are at (0,0,0) and Pb atoms
are at (0,0,1/2) positions.}
\end{figure}

The electrical resistivity $\rho$ was measured from 0.35 K to 300 K.
As shown in Fig.~2,  the resistivity is that of a good metal,
decreasing from the value of 16 $\mu\Omega\cdot$cm at 300 K to 2
$\mu\Omega\cdot$cm at the lowest temperature. The inset to Fig.~2
shows a partial superconducting transition at T$_{c}$=7.2 K due to
residual Pb flux, followed by another partial superconducting
transition at T$_{c}$=3 K, which we believe is due to trace amounts
of superconducting RhPb$_{2}$~\cite{gendron1962}. We performed
Meissner effect measurements which confirmed these conclusions,
finding a volume fraction of less than 1$\%$ for the proposed Pb
inclusions, and an even smaller volume fraction for RhPb$_{2}$. The
almost linear temperature dependence of $\rho$(T) demonstrated in
Fig.~2 for temperatures above $\sim$ 70 K is remarkable. While this
linearity is manifestly not that of a normal Fermi liquid ground
state, it may indicate that the resistivity is derived from the
electron-phonon interaction and further that the Debye temperature
in YbRh$_{2}$Pb is unusually small. However, we later show, using
the heat capacity data that the Debye temperature in YbRh$_{2}$Pb is
$\sim$213 K. It is notable as well that the resistivity does not
saturate at high temperatures, suggesting that the electronic mean
free path is still much larger than the lattice constant at room
temperature. We cannot entirely rule out the possibility that the
fundamental electronic excitations are themselves anomalous, as was
found in the normal state of the high temperature oxide
superconductors.~\cite{civale1988}

\begin{figure}
\includegraphics[scale=0.55]{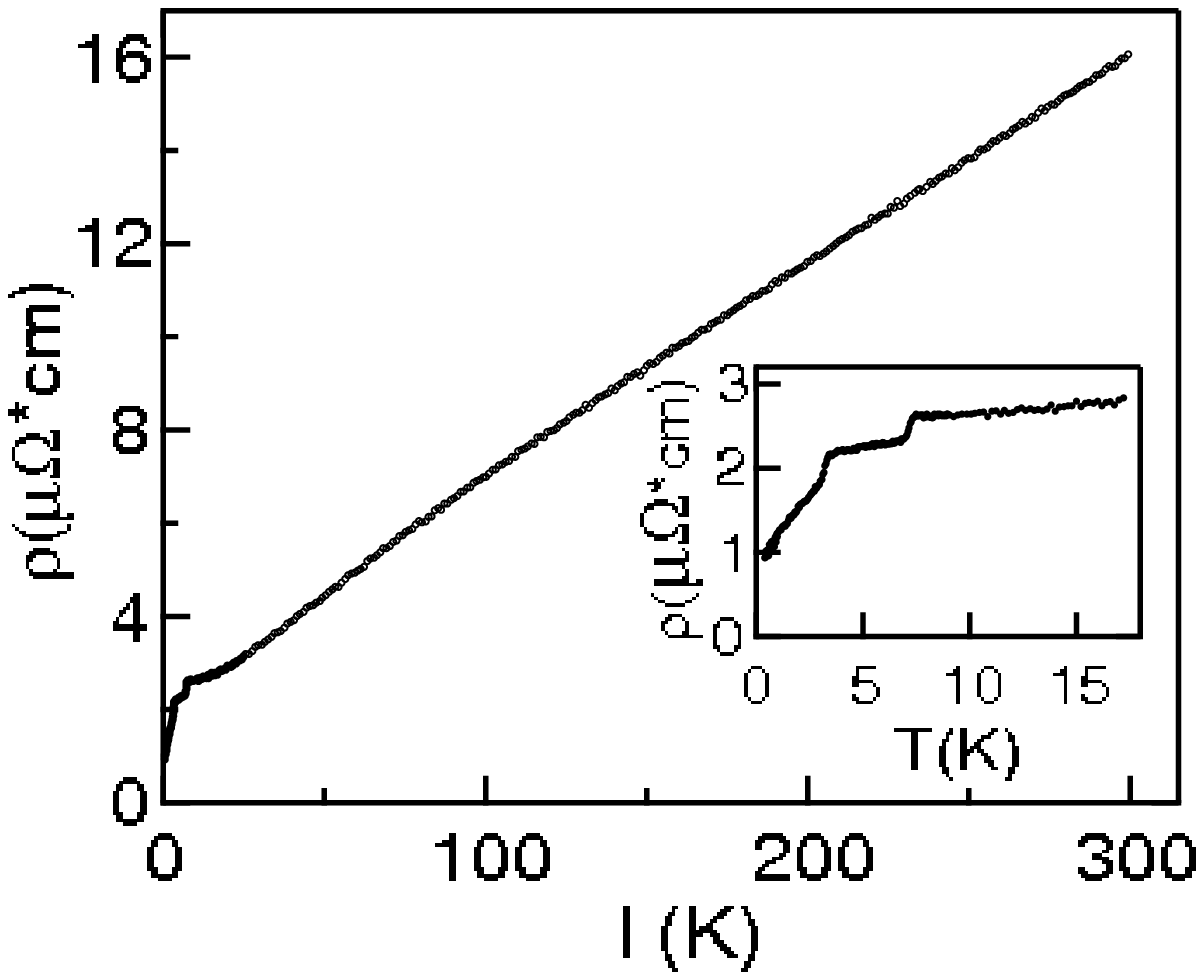}
\caption{\label{fig:epsart} Temperature dependence of the electrical
resistivity of YbRh$_{2}$Pb measured in zero field. Inset: Expanded
view of low temperatures, showing two partial superconducting
transitions at 7.2 K and 3 K due to trace amounts of Pb and
RhPb$_{2}$, respectively.}
\end{figure}

The magnetic properties of YbRh$_{2}$Pb establish that the Yb$^{3+}$
moments are well localized. The temperature dependence of the
$\emph{dc}$ magnetic susceptibility $\chi$(T) was measured in a
field of 1000 Oe. The data are plotted as 1/$\chi$(T) as a function
of temperature in Fig.~3. These data are only linear between 100 K
and 300 K, where the Curie-Weiss analysis gives a Weiss temperature
$\Theta$=-1.9$\pm$0.1 K and an effective magnetic moment
$\mu_{eff}$=3.3$\pm$0.1 $\mu_{B}$ per Yb ion, a substantial fraction
of the value of 4.54 $\mu_{B}$ expected for a free Yb$^{3+}$ ion.
The inset shows that 1/$\chi$ deviates below the linear behavior at
T$<$20 K, but remains finite at the lowest temperature. We will
argue below that the general lack of agreement between the measured
susceptibility and a single Curie-Weiss expression results from
significant CEF effects, yielding a highly temperature dependent
effective moment in YbRh$_{2}$Pb.

\begin{figure}
\includegraphics[scale=0.55]{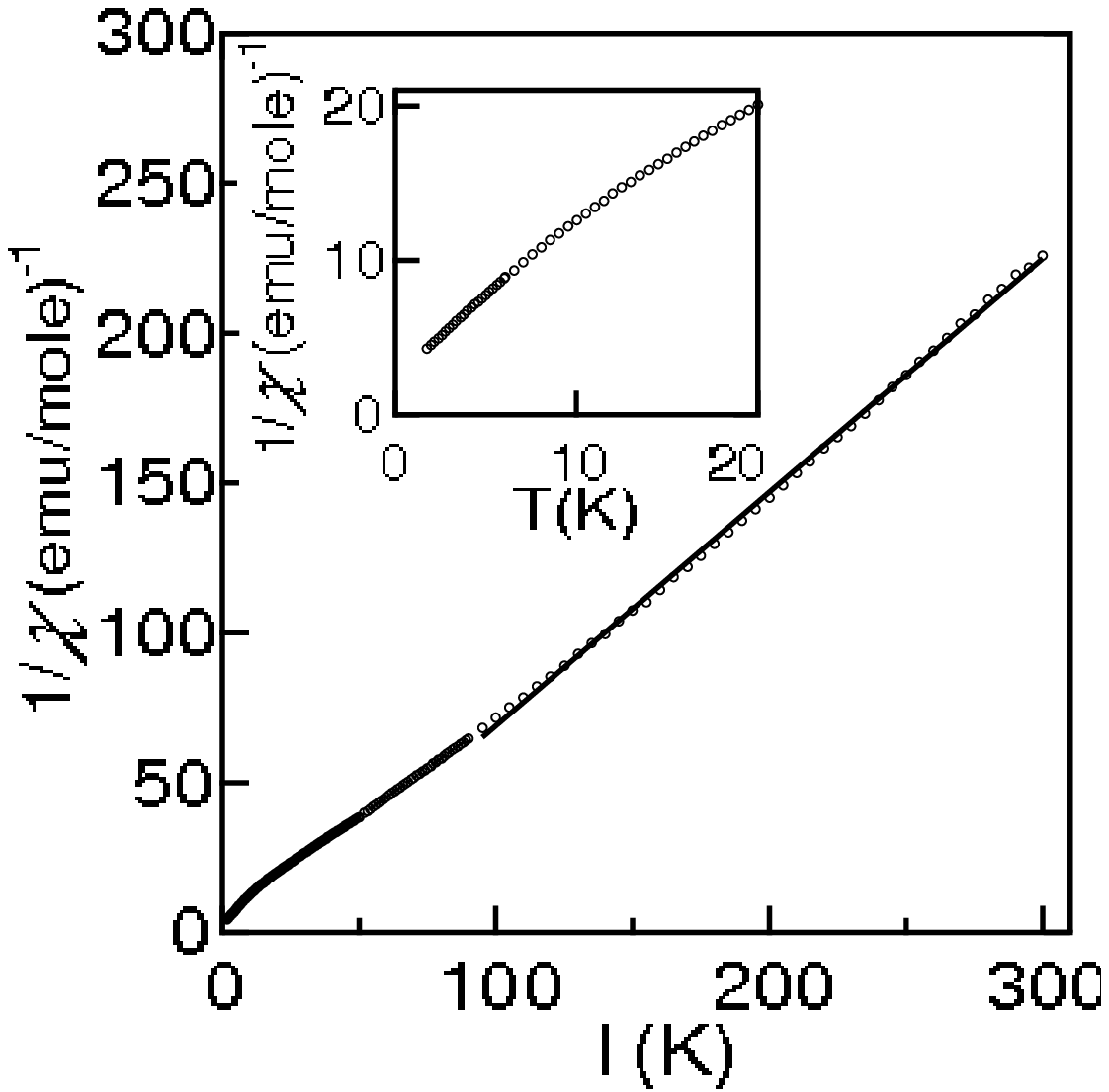}
\caption{\label{fig:epsart} Temperature dependence of the inverse of
the magnetic susceptibility $\chi^{-1}$ of YbRh$_{2}$Pb measured in
a field of 1000 Oe. Solid line is linear fit to
1/$\chi$=T/C-$\Theta$/C for temperatures between 100 K and 300 K,
where C is the Curie constant and $\Theta$ is the Weiss temperature.
The inset shows 1/$\chi$ below T=20 K.}
\end{figure}

We derived the basic ingredients for devising a crystal field scheme
for YbRh$_{2}$Pb from heat capacity measurements. The temperature
dependence of the heat capacity C(T) measured in zero field and at
temperatures as large as 70 K is shown in Fig.~4a. We have estimated
the phonon contribution to the heat capacity C$_{ph}$ using the
Debye expression and find a Debye temperature $\theta_{D}$=213$\pm$5
K. C$_{ph}$ is subtracted from C(T) in Fig.~4a to isolate the
remaining magnetic and electronic contributions to the heat
capacity. The latter is expected to result in a component of the
heat capacity which is linear in temperature, C$_{el}$=$\gamma$T.
Accordingly we have plotted C/T as a function of T$^{2}$ in Fig.~4b,
demonstrating that the electronic contribution can at best be
identified over a very limited range of temperatures, yielding
$\gamma$=4$\pm$1 mJ/moleK$^{2}$. We conclude that the purely
electronic contribution to the heat capacity is very small, as would
be expected for weakly correlated conduction electrons or
alternatively for a low density of conduction electrons for
YbRh$_{2}$Pb. This implies, in turn, that C-C$_{ph}$ is largely
magnetic. An expanded view of the temperature dependence of
C-C$_{ph}$ is presented in Fig.~5, accompanied by the associated
entropy S. The sudden increase in C-C$_{ph}$ near 0.57 K indicates a
magnetic phase transition, possibly superposed on a broad Schottky
peak. We have approximated C-C$_{ph}$ to zero temperature to account
for the part of the heat capacity below 0.35 K inaccessible in our
measurement. The entropy difference associated with the magnetic
phase transition is only 0.78 Rln2, indicating that magnetic order
does not develop from a well defined doublet ground state,
alternatively an onset of the weak Kondo effect with the Kondo
temperature T$_{K}$$<$0.5 K can account for a reduced entropy. The
entropy is roughly constant between 2 K and 10 K, subsequently
increasing in proximity to a large Schottky anomaly centered at 28 K
and approaching the full Rln2 only near 30 K.

\begin{figure}
\includegraphics[scale=0.55]{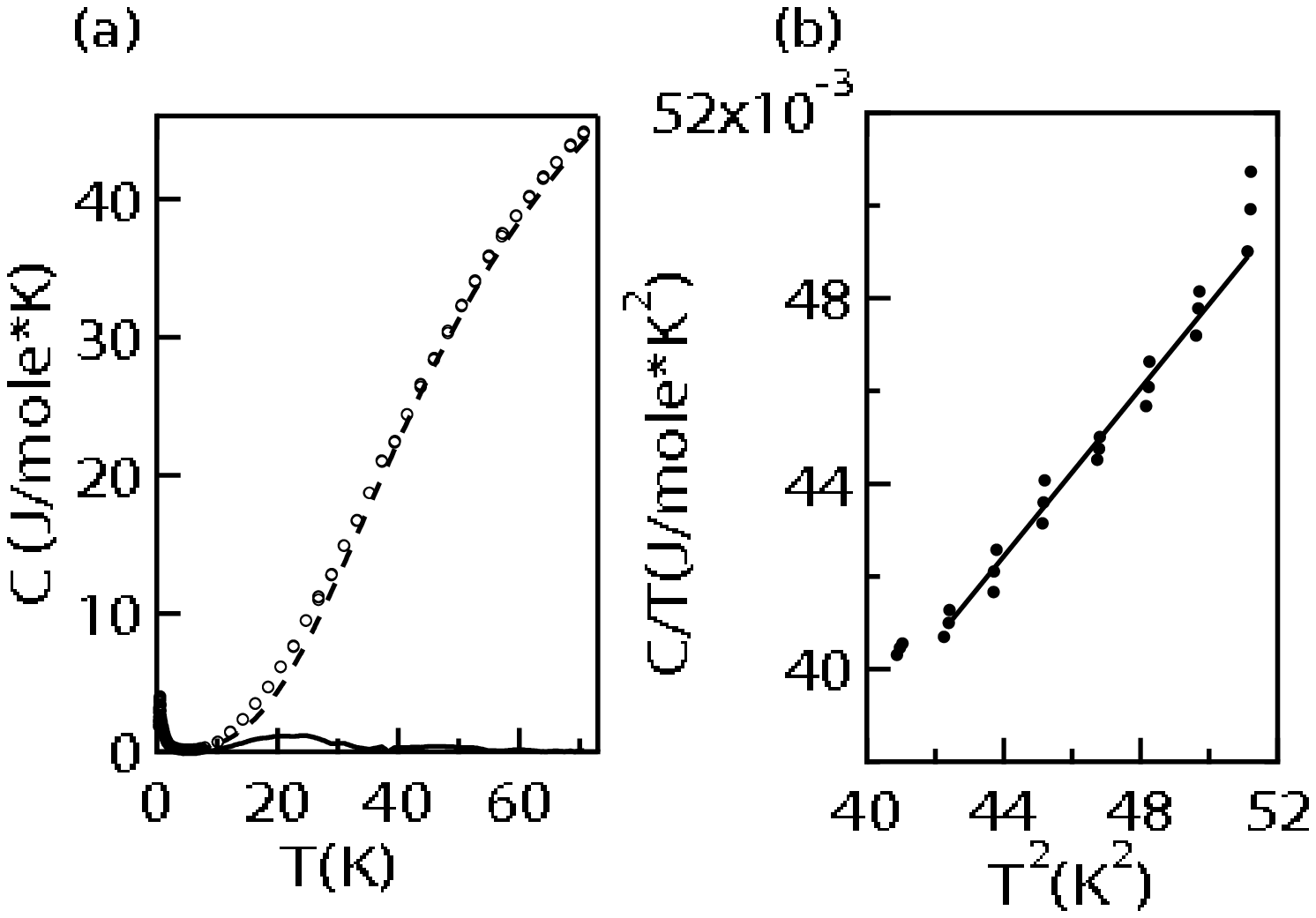}
\caption{\label{fig:epsart} (a) Temperature dependence of the heat
capacity C ($\circ$) of YbRh$_{2}$Pb measured in zero field. The
dashed line represents the Debye heat capacity, the solid line is
the heat capacity with the lattice contribution subtracted. (b) The
electronic part of the heat capacity C$_{el}$=$\gamma$T is
determined from this plot of C/T=$\gamma$+$\beta$T$^2$.}
\end{figure}

\begin{figure}
\includegraphics[scale=0.55]{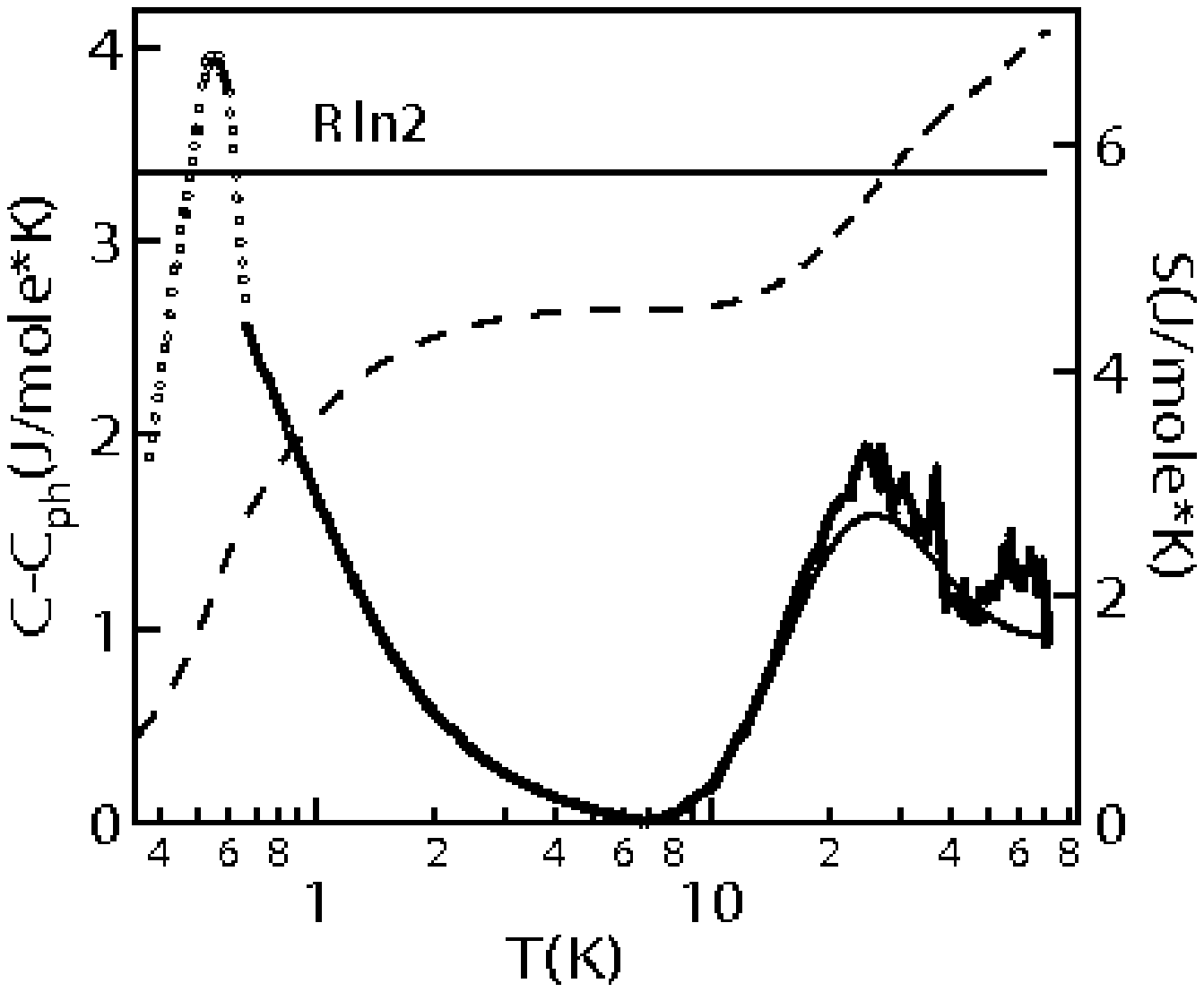}
\caption{\label{fig:epsart} Temperature dependence of the non-phonon
heat capacity C-C$_{ph}$ ($\circ$) and the entropy S (dashed line)
of YbRh$_{2}$Pb measured in zero field. The peak in C-C$_{ph}$
centered at 28 K is well fit by the Schottky expression described in
the text (solid line).}
\end{figure}

Although the site symmetry of Yb in YbRh$_{2}$Pb is tetragonal and
hence the CEF scheme must consist of 4 doublets, we show that the
first excited doublet lies sufficiently close to the ground state
doublet that the ground state can be considered a quartet. In Fig.~6
we schematically illustrate the CEF splitting scheme in
YbRh$_{2}$Pb, which is derived from our heat capacity measurements.
The best fit to the broad maximum in the heat capacity at $\sim$28 K
(Fig.~5) was obtained assuming a fourfold degenerate ground state
separated by an energy of $\Delta_{1}$=68$\pm$5 K  from the first
excited state, which is a doublet, and from the second excited
state, also a doublet, by the energy $\Delta_{2}$=300$\pm$60 K.  We
conclude that the full Yb moment can only be regained at
temperatures much higher than 300 K, and that at lower temperatures
there is a substantially smaller and temperature dependent effective
Yb moment.

\begin{figure}
\includegraphics[scale=0.7]{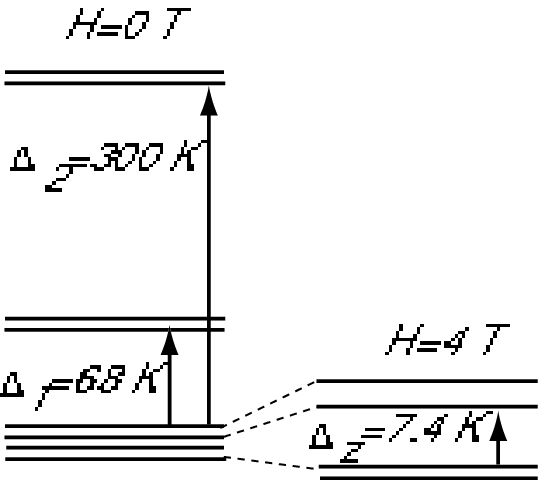}
\caption{\label{fig:epsart} Proposed zero field CEF scheme for
Yb$^{3+}$ in YbRh$_{2}$Pb (left). Application of a magnetic field
lifts the degeneracy of the ground state quartet by an amount
$\Delta_{z}$, which is 7.4 K at 4 T (right).}
\end{figure}

\begin{figure}
\includegraphics[scale=0.6]{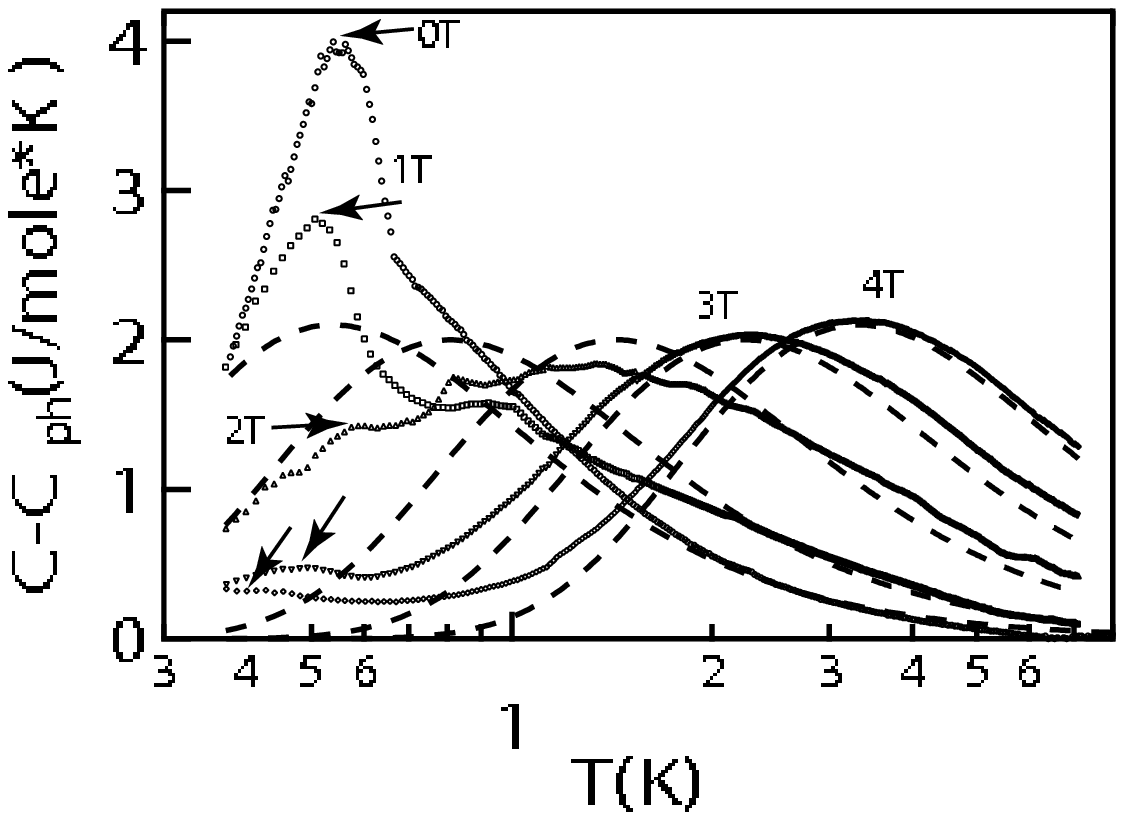}
\caption{\label{fig:epsart} Temperature dependence of the heat
capacity of YbRh$_{2}$Pb measured in magnetic fields with the
lattice contribution subtracted. Dashed lines represent the Schottky
heat capacity C$_{sch}$ of the two-level system, when the levels are
split by the magnetic field.}
\end{figure}

To further test this CEF scheme we have measured C(T) in a variety
of fields between zero and 4 T. The data are presented in Fig.~7.
Magnetic fields suppress the heat capacity jump at the magnetic
transition. The ordering temperature also decreases in field, and
Fig.~8 indicates that a magnetic field $>$4 T is required to
completely suppress magnetic order. For fields greater than $\sim$2
T, the measured heat capacity mostly consists of a single broad peak
which shifts to higher temperatures with increasing magnetic field.
We attribute this peak to the Zeeman splitting of the ground state
of YbRh$_{2}$Pb, which is a quartet in zero field. The field
dependence of the heat capacity represented in Fig.~7 can be
understood by assuming that the quartet is Zeeman split by the field
into a ground state doublet and two excited singlets, as depicted in
Fig.~6. Accordingly, in a field of 4 T, the heat capacity reaches
$\sim$2 J/mole$\cdot$K at T$_{max}$=0.448$\Delta_{z}$=3.3 K as
expected for a system with a thermally activated occupation levels
with degeneracy ratio of 0.5~\cite{sereni1991}. The splitting
$\Delta_{z}$ between the ground doublet and first excited doublet is
deduced from Schottky fits to the data of Fig.~7, and $\Delta_{z}$
is itself plotted in Fig.~8. We note that including a second excited
singlet state to this analysis did not significantly improve the
quality of the fits. $\Delta_{z}$ increases approximately linearly
to a value of 7 K in a field of 4 T. Interestingly, Fig.~8 suggests
that there is a residual splitting of $\sim$ 1.5 K in zero field,
which may perhaps account for the broad peak on which the magnetic
anomaly is superposed (Fig.~5). Apart from this small splitting, the
evolution of the heat capacity with magnetic fields indicates that
the zero field ground state of YbRh$_{2}$Pb is fourfold degenerate.
It would be interesting to further test this level scheme in
YbRh$_{2}$Pb using inelastic neutron scattering experiments.

\begin{figure}
\includegraphics[scale=0.45]{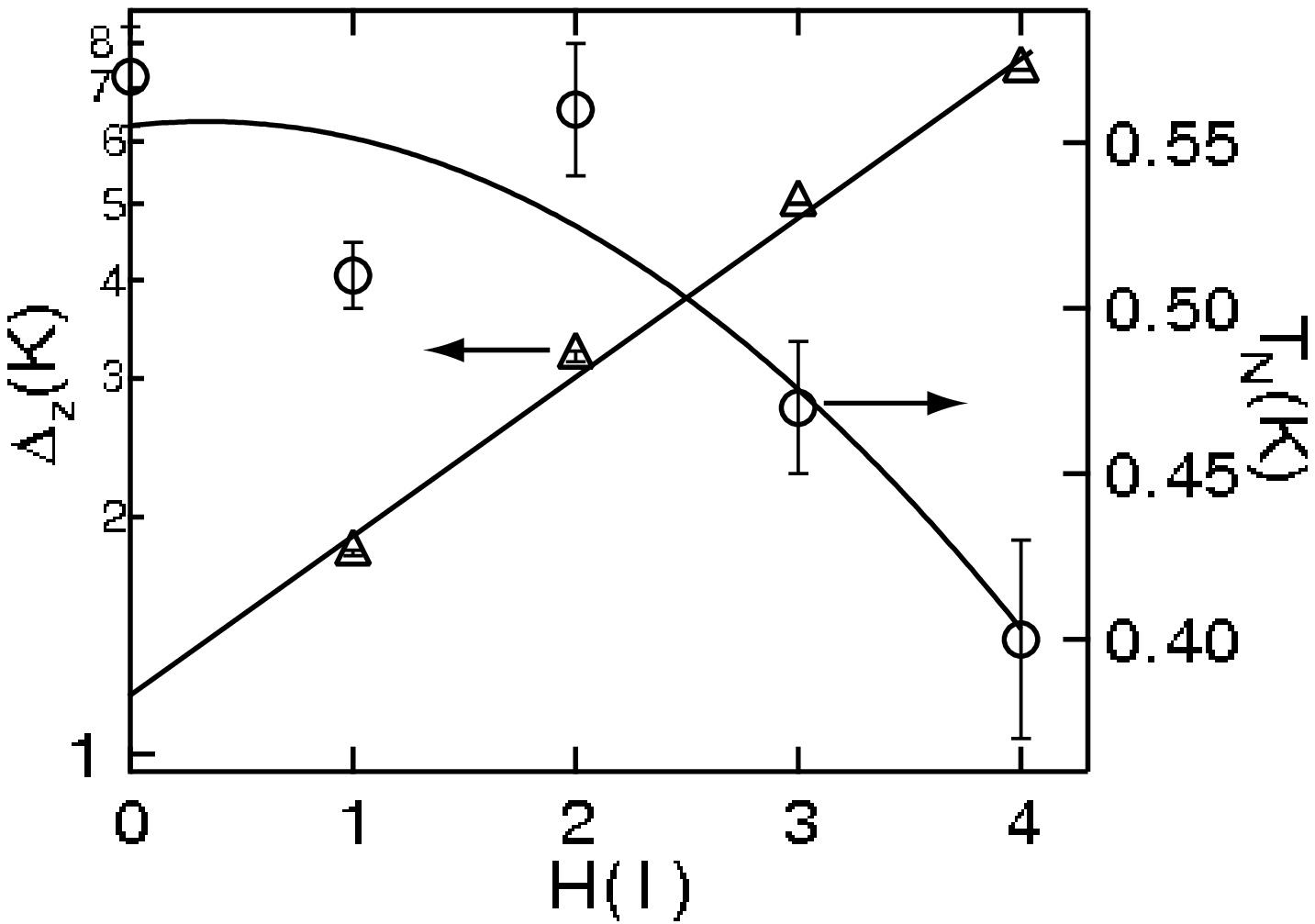}
\caption{\label{fig:epsart} Suppression of the magnetic ordering
temperature T$_{N}$ with increasing magnetic field ($\circ$);
$\bigtriangleup$ indicates the  Zeeman splitting $\Delta_{z}$ of the
ground state quartet, taken from the Schottky analysis of the heat
capacity described in the text. Solid lines are guides for the eye.}
\end{figure}

Given the magnitudes of the splittings in the CEF scheme of
YbRh$_{2}$Pb which are revealed by our analysis of the heat
capacity, we expect that the effective Yb moment varies considerably
over the temperature range 1.8 K - 300 K spanned by our
measurements. As we have observed in our discussion of Fig.~3, this
implies that $\chi$ cannot be fit by the Curie-Weiss law over an
extensive range of temperatures. This is illustrated in the Fig.~9,
where we plotted the effective magnetic moment $\mu_{eff}
(T)=\sqrt{\frac{3k_{B}T\chi}{N}}$ as a function of temperature. With
increased temperature, $\mu_{eff}$ increases as the excited states
become occupied. However, the full Yb$^{3+}$ Hund's rule moment of
4.54 $\mu_{B}$ is not regained even at temperatures as large as 300
K.

The interplay between moment degeneracy and intermoment interactions
strongly impacts the stability of magnetic order in YbRh$_{2}$Pb,
which only occurs below 0.57 K. While neutron diffraction
measurements would be required to unambiguously establish the type
of magnetic order, the relatively weak field dependence of the
transition temperature and the absence of appreciable broadening of
the transition in the heat capacity suggest that YbRh$_{2}$Pb is an
antiferromagnet. Magnetic order arises within a ground state
quartet, which is further split into a doublet separated from the
first of two excited singlets by $\sim$ 1.5 K. Increasing this
separation $\Delta_{z}$ with magnetic field results in the further
suppression of the ordering temperature T$_{N}$.  We observe that
the successive actions of tetragonal and uniaxial CEF effects
results in a paramagnetic state with a moment much reduced from the
free ion value. At the same time, the small residual resistivity and
the small electronic heat capacity suggest extremely weak
hybridization between the Yb moment and the conduction electrons,
and subsequently a small RKKY interaction among Yb moments. For
these reasons, we believe that magnetic order in YbRh$_{2}$Pb is
driven by a vanishingly small RKKY interaction acting on moments
which have been reduced to a minimal value by CEF splittings which,
while small in an absolute sense, are very large compared to T$_{N}$
itself.

\begin{figure}
\includegraphics[scale=0.5]{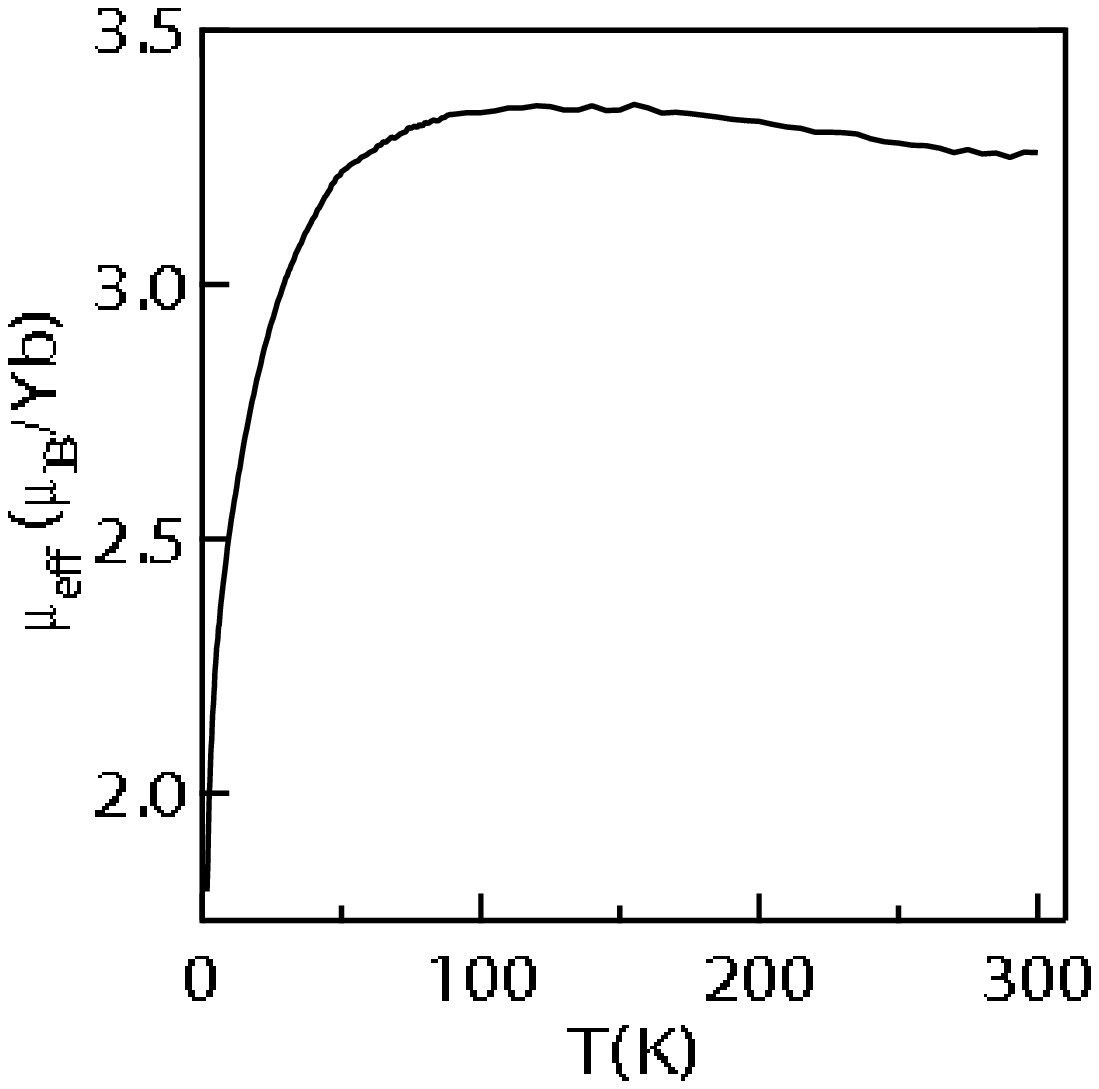}
\caption{\label{fig:epsart} The temperature dependent effective
magnetic moment $\mu_{eff} (T)=\sqrt{\frac{3k_{B}T\chi}{N}}$ vs T.}
\end{figure}

\section{CONCLUSION}
We have synthesized the novel intermetallic Heusler compound
YbRh$_{2}$Pb, which orders magnetically below T$_{N}$=0.57 K. The
electrical resistivity displays an unusual linear temperature
dependence, and its small residual value is consistent with good
crystalline quality. Analyses of the heat capacity and the magnetic
susceptibility indicate that the CEF splits the eight-fold
degenerate ground state of Yb$^{3+}$ into a ground state quartet and
two excited doublets separated from the ground state by 68 K and 300
K, respectively. Magnetic fields further split the ground state, and
we propose that this reduction in the effective moment is
responsible for the observed reduction in the ordering temperature.
Given the strong current interest in the quantum critical behavior
which is found near phase transitions which occur at low or zero
temperatures~\cite{stewart2001}, it is interesting to consider
whether YbRh$_{2}$Pb belongs to this class of materials. While
YbRh$_{2}$Pb orders at a very low temperature, one which can be
reduced still further by applying a magnetic field, it is unlikely
that quantum critical behavior will be observable in this system.
The purely local moment nature of the magnetism, paired with the
absence of correlation effects among the conduction electrons
suggest that YbRh$_{2}$Pb belongs instead to the weak coupling limit
of the Doniach phase diagram~\cite{doniach1977}. Without a large
magnetic energy scale, the conventional critical phenomena
associated with the small but finite temperature magnetic phase
transition will limit any non-Fermi liquid behavior in YbRh$_{2}$Pb
to very small reduced temperatures~\cite{millis1993}.

\section{ACKNOWLEDGMENTS}
Work at the University of Michigan was performed under grant
NSF-DMR-0405961. D. A. S. acknowledges useful conversations with Z.
Fisk. The electron microprobe used in this study was partially
funded by grant EAR-99-11352 from the National Science Foundation.
Use of the National Synchrotron Light Source, Brookhaven National
Laboratory, was supported by the U.S. Department of Energy, Office
of Science, Office of Basic Energy Sciences, under Contract No.
DE-AC02-98CH10886. We are grateful to Joseph Lauher for the use of
X-ray diffraction facilities at the Stony Brook University Chemistry
Department.

$\dag$ e-mail address: sokolov@bnl.gov (D. A. Sokolov)

\end{document}